\documentclass[AMA,Times2COL]{neutral}
\usepackage{graphicx}
\usepackage[]{xcolor}
\usepackage{xspace}
\usepackage{listofitems}
\usepackage{booktabs}
\usepackage{multirow}
\usepackage{subcaption}

\newcommand{\qb}{\mathbf{q}}
\newcommand{\qq}{\mathbf{q}}

\newcommand{\qnu}{{{\qb\nu}}}
\newcommand{\wqnu}{{\omega_{\qnu}}}

\newcommand{\abinit}{\textsc{Abinit}\xspace}
\newcommand{\anaddb}{\texttt{anaddb}\xspace}

\newcommand{\abipy}{AbiPy\xspace}

\newcommand{\phonopy}{\textsc{phonopy}\xspace}

\newcommand{\abinitio}{\textit{ab initio}\xspace}

\newcommand{\ie}{{\emph{i.e.}}}
\newcommand{\eg}{{\emph{e.g.}}}

\newcommand{\Zstar}{{\bf Z}^*}

\newcommand{\problematic}{\textit{problematic}\xspace}
\newcommand{\unconverged}{\textit{unconverged}\xspace}
\newcommand{\converged}{\textit{converged}\xspace}

\articletype{Research Article}
\received{Date Month Year}
\revised{Date Month Year}
\accepted{Date Month Year}
\journal{Journal}
\volume{00}
\copyyear{2024}
\startpage{1}

\begin{document}
\title{Systematic assessment of various universal machine-learning interatomic potentials}

\author[1]{Haochen Yu}
\author[1]{Matteo Giantomassi}
\author[1]{Giuliana Materzanini}
\author[2]{Junjie Wang}
\author[1,2,3]{Gian-Marco Rignanese}

\authormark{YU \textsc{et al.}}
\titlemark{SYSTEMATIC ASSESSMENT OF VARIOUS UNIVERSAL MACHINE-LEARNING INTERATOMIC POTENTIALS}

\address[1]{\orgdiv{Institute of Condensed Matter and Nanosciences}, \orgname{Université catholique de Louvain}, \orgaddress{1348 Louvain-la-Neuve, \country{Belgium}}}
\address[2]{\orgdiv{State Key Laboratory of Solidification Processing}, \orgname{Northwestern Polytechnical University}, \orgaddress{Xi’an, Shaanxi 710072, \country{Republic of China}}}
\address[3]{\orgname{WEL Research Institute}, \orgaddress{1300 Wavre, \country{Belgium}}}

\corres{Corresponding author Gian-Marco Rignanese, \email{gian-marco.rignanese@uclouvain.be}}

\abstract[Abstract]{
Machine-learning interatomic potentials have revolutionized materials modeling
at the atomic scale.
Thanks to these, it is now indeed possible to perform simulations of \abinitio
quality over very large time and length scales.
More recently, various universal machine-learning models have been proposed as
an out-of-box approach avoiding the need to train and validate specific
potentials for each particular material of interest.
In this paper, we review and evaluate four different universal machine-learning
interatomic potentials (uMLIPs), all based on graph neural network architectures
which have demonstrated transferability from one chemical system to another.
The evaluation procedure relies on data both from a recent verification study of
density-functional-theory implementations and from the Materials Project.
Through this comprehensive evaluation, we aim to provide guidance to materials
scientists in selecting suitable models for their specific research problems,
offer recommendations for model selection and optimization, and stimulate
discussion on potential areas for improvement in current machine-learning
methodologies in materials science.
}


\maketitle
\section{Introduction}
Materials simulations at the atomic scale are the backbone of computational
materials design and discovery.
They rely on the Born Oppenheimer approximation, in which the electrons follow
the nuclear motion adiabatically, so that the potential governing the nuclei
consists of the electronic energies as a function of the nuclear positions,
called "potential energy surface" (PES).
Knowledge of the PES allows the identification of stable and metastable atomic
configurations from minimum energy search, or the determination of materials
properties as thermodynamical averages from molecular dynamics
simulations~\cite{frenkel2023understanding}.
The utility of atom-based materials simulations is thus intimately related to
the generation of accurate PESs, which has been possible in the last decades
thanks to the advent of density-functional theory (DFT)~\cite{Hohenberg1964,
Kohn1965,payne1992iterative,Marzari2021}.
Nonetheless, this \abinitio approach relies on the quantum mechanical solution
of the electronic problem whose computational cost scales
cubically with system size and can therefore become unaffordable in various
significant cases of technological interest such as amorphous solids,
interfaces, surfaces, etc.
At the other end of the simulation approaches, parametrized approximations of
the Born-Oppenheimer PES, known as empirical analytical potentials, or force
fields, or "classical” interatomic potentials, have been widely used especially
for large-scale materials studies~\cite{Goddard2021}.
Unfortunately, in particular when complex electron interactions are involved (as
in chemical reactions or phase transitions) these approaches cannot usually
achieve DFT accuracy, and in addition they have limited applicability and
transferability.
They cannot therefore be considered as a drop-in replacement for standard
\abinitio methods.
In this context, machine-learning interatomic potentials (MLIPs) have emerged as
an in-between solution with computational cost similar to the empirical
analytical potentials, but with the promise of achieving an accuracy comparable
to DFT hence enabling accurate simulations over very large time and length
scales~\cite{zuo2020performance}.
The key difference with respect to the empirical potentials is that the
interatomic potentials are now directly obtained via a highly non linear fit of
a set of input/target data (in general, at the DFT accuracy), without any a
priori assumption on their analytical form~\cite{Deringer2019}.
In the original formulation, building a MLIP consists of generating a dataset of
atomic configurations for the specific material under study, and training (and
subsequently validating) the MLIP on these data based on the accurate prediction
of some target metrics as, \eg, energies, forces, and stresses~\cite{
Deringer2019, Behler2007, Bartok2010, Thompson2015, Schutt2018, Drautz2019,
vonLilienfeld2020, Batzner2022, Ko2023}.
This process is highly material-dependent, and usually requires a significant
human and computational effort~\cite{Deringer2021}.
Obtaining an accurate description of the PES without the need for costly DFT
computations, but also covering all possible chemical and structural spaces,
would be the holy grail of MLIPs.
In this framework, graph Neural Networks (GNN) have revealed very effective, in
particular in message passing and equivariant~\cite{Batzner2022,geiger2022e3nn}
NN~\cite{Batatia2022}.
In general, NN-based methods can be particularly useful for generalizability
thanks to their property of “learning locally” that makes the resulting
potential less material-dependent~\cite{,Deringer2019,
Behler2007,behler2011neural,behler2017first,behler2021four}.
Indeed, the first ``universal'' MLIP (uMLIP), the MEGNet~\cite{Chen2019} model,
exploited a graph network architecture.
It was trained on $\sim$60000 inorganic crystals in their minimum energy
configurations from the Materials Project (MP)~\cite{Jain2013} database, which
covers 89 elements of the periodic table and is based on the
Perdew-Burke-Enzerhof (PBE) exchange-correlation functional~\cite{Perdew1996}.
This model could provide the formation energy as well as a number of other
properties.
In order to predict forces and stresses, various other models were subsequently
developed relying on different datasets.
In particular, the M3GNet~\cite{Chen2022} and CHGNet~\cite{Deng2023} models
relying on equivariant graph neural network architectures were both developed
using snapshots from DFT relaxations of the MP structures.
The publication introducing CHGNet was also the opportunity for releasing the
Materials Project Trajectory (MPtrj) dataset~\cite{Deng2023}, including the DFT
calculations for more than 1.5 million atomic configurations of inorganic
structures.
By including magnetic moments in the training properties, CHGNet aimed at a
better description of chemical reactions as charged states influence how atoms
connect with others through chemical bonds~\cite{Deng2023}.
A re-implementation of M3GNet, MatGL, has been built on the Deep Graph Library
and on PyTorch in order to improve usability and
scalability~\cite{matgl-github}.
Recently, the M3GNet-DIRECT model has been released~\cite{qi2024robust} in the
MatGL package~\cite{matgl-github} with 1.3 million structures and 89 elements,
aiming to predict more reliably unseen structures thanks to a new sampling
scheme~\cite{qi2024robust}.
The ALIGNN-FF model~\cite{Choudhary2023} was developed to model a diverse set of
materials with any combination of 89 elements from the periodic table but
relying on a different database of inorganic crystals,
JARVIS-DFT~\cite{Choudhary2020}, which is based on the optB88vdW
exchange-correlation functional~\cite{Klimes2009}.
Recently, the MACE-MP-0~\cite{batatia2023foundation} relying on the MACE
architecture~\cite{Batatia2022} was also proposed.
It was trained on the MPtrj dataset and showed outstanding performance on an
extraordinary range of examples from quantum-chemistry and materials
science~\cite{batatia2023foundation}.
Two proprietary models relying on very large databases have also been developed:
the GNoME~\cite{Merchant2023} model exploiting the NequIP
architecture~\cite{Batzner2022} and the PFP model~\cite{Takamoto2022,
Takamoto2023} exploiting the TeaNet architecture~\cite{Takamoto2022a}.
On the one hand, GNoME was trained on a database obtained from a complex active
learning workflow of the original MP data, resulting in a number of inorganic
structures $\sim$100 times larger than the MPtrj.
On the other hand, the PFP was trained on a large dataset which initially
included $\sim$10$^7$ DFT conﬁgurations covering 45 elements~\cite{Takamoto2022}
and was further extended to cover 72 elements at the moment of the
publication~\cite{Takamoto2023}, with an expected rise to 94 including
rare-earth elements and actinides.
It is worth mentioning that uMLIPs have also been developed specifically for
organic molecules~\cite{Smith2017, Smith2019, Zubatyuk2019}, and for metal
alloys~\cite{Lopanitsyna2023}.
While this rapid growth of uMLIP models promises an extraordinary impact in the
materials science community, nevertheless benchmark efforts are crucially needed
to assess and control their performance and effective usefulness
\cite{riebesell2023matbench,focassio2024performance,deng2024overcoming}.

In this paper, we conduct a comprehensive review and evaluation of four
different GNN-based uMLIPs: M3GNet-DIRECT~\cite{qi2024robust}, 
CHGNet~\cite{Deng2023}, MACE-MP-0~\cite{batatia2023foundation}, and
ALIGNN-FF~\cite{Choudhary2023}.
They have demonstrated the possibility of universal interatomic potentials that
may not require retraining for new applications. 
The evaluation uses three different datasets: for the equation of state, we use
the set of theoretical structures employed in Ref.~\cite{Bosoni2023}, for the
phonon calculations, we use the crystalline structures considered in
Ref.~\cite{Petretto2018a} that have been relaxed with
\abinit~\cite{Gonze2020,Romero2020}, norm-conserving
pseudopotentials~\cite{Hamann2013, VanSetten2018}, and the PBEsol
exchange-correlation functional~\cite{perdew2008restoring} while for all the
other tests, we use the VASP-relaxed structures from the Materials Project
(MP)~\cite{Jain2013,Kresse1999}.

The paper is organized as follows.
After a brief overview of the techniques employed in the generation of the
different uMLIPs examined in this work (Sec.~\ref{Subsec:potentials_details}),
we test and discuss the quality and transferability of the chosen uMLIPs in
different types of calculations, and using the three different datasets as
explained above.
The tests include the calculation of the equation of state
(Sec.~\ref{Sec:common_workflow}), the evaluation of formation energies and 
optimization of structural parameters (Sec.~\ref{Sec:relaxations}), and the
calculation of phonon bands (Sec.~\ref{Sec:phonons}).
The conclusions are provided in Sec.~\ref{Sec:conclusions} and the details of
the methods employed are given in Sec.~\ref{Sec:methods}.

\section{Discussion}\label{Sec:discussion}

\subsection{Overview of the uMLIP models}
\label{Subsec:potentials_details}

In what follows, we give a brief overview of the basic ingredients underlying
the four GNN-based uMLIPs
examined~\cite{Deng2023,qi2024robust,Choudhary2023,batatia2023foundation}.
Depending on the uMLIPs, they might differ for the specific GNN architecture
employed, for the choice of the training set, for the sampling scheme, or for
the properties included in the training. A comprehensive and detailed
explanation is beyond the scopes of the present work, and we refer the
interested reader to the respective original
papers~\cite{Deng2023,qi2024robust,Choudhary2023,batatia2023foundation} and
related literature~\cite{Batatia2022,Chen2019,Chen2022,Choudhary2020,
Chen2021,Choudhary2022,batatia2022design,Choudhary2021}. 
GNN are reviewed, among others, in
Refs.~\cite{geiger2022e3nn,wu2020comprehensive,asif2021graph}, while
Refs.~\cite{ Deringer2019, Behler2007, Bartok2010, Thompson2015,
Schutt2018,Drautz2019, vonLilienfeld2020, Batzner2022, Ko2023,
Deringer2021,mueller2020machine} address specifically the construction of MLIP
in materials science.
\\
The pre-trained M3GNet-DIRECT~\cite{qi2024robust} model starts from the Materials 3-body Graph Network (M3GNet) architecture~\cite{Chen2022}, which conjugates the flexibility of typical graph techniques with the physically based many-body features (here 3-body) of traditional interatomic potentials like the Tersoff bond-order potential~\cite{tersoff1988new}.
The many-body interactions (angles) are aggregated to bonds in standard graph convolution steps to update the bond, atom, and global state (properties) information~\cite{Chen2022}.
The main improvement of M3GNet-DIRECT~\cite{qi2024robust} with respect to the original M3GNet~\cite{Chen2022} is in the sampling of the materials space. This is done starting from the 128-element vector outputs from the M3GNet~\cite{Chen2022} model trained on the MP formation energies database~\cite{Jain2013}, to which dimensionality reduction, characteristics-sharing-based clustering, and stratification, are subsequently applied, giving the name of DImensionality-Reduced Encoded Clusters with sTratified (DIRECT) sampling to the approach~\cite{qi2024robust}. Static DFT calculations on the so-obtained 1.3 million structures produce the final training set~\cite{qi2024robust}. The result is an improved performance with respect to the original M3GNet~\cite{Chen2022}, with a better transferability and prediction of configurations with large energies, forces, and stresses~\cite{qi2024robust}.
\\
The Crystal Hamiltonian Graph Neural Network (CHGNet~\cite{Deng2023}) is a model pre-trained on the energies, forces, stresses, and magnetic moments from the MPtrj~\cite{Deng2023} dataset. The GGA/GGA+U mixing compatibility corrections~\cite{Wang2021,Jain2011,U-corrections} are applied to the energies~\cite{Deng2023}.
In the CHGNet architecture, the angle information is drawn as a pairwise message passing convolution between bonds (bond graph, where bonds are nodes and edges are angles), on top of the bond information which is drawn as a pairwise convolution between atoms (atom graph, where atoms are nodes and edges are bonds)~\cite{Deng2023}. 
To constrain the atom features used to predict energy, forces, and stresses by their charge-state
information, the latter is inferred from the magnetic moments and atomic orbital theory before the last convolution layer~\cite{Deng2023}.
In incorporating the site-specific magnetic moments as the input charge states into the GNN, this model aims to capture the chemical interaction variability across different valence states~\cite{Deng2023}. This should be especially important for transition-metal elements, with variable valence states~\cite{reed2004role}.
\\
The pre-trained MACE-MP-0 model~\cite{batatia2023foundation} relies on MACE~\cite{Batatia2022}, an equivariant message passing neural network potential using messages of an order higher than two body. In combining high body order with message passing, MACE achieves superior accuracy as a result of a trade-off between using more layers and increasing the local body order within a single layer~\cite{batatia2022design}.
Using 4-body messages~\cite{Batatia2022}, the required number of message passing iterations is reduced to two and the steepness of the learning curve is improved~\cite{Batatia2022}. 
The high-order equivariant message passing neural network model is combined with the atomic cluster expansion (ACE~\cite{Drautz2019, dusson2022atomic}), a framework for deriving an efficient body-ordered symmetric polynomial basis
to represent functions of atomic neighborhoods. 
ACE naturally extends to
equivariant features and to include variables beyond geometry, such as charges or magnetic moments~\cite{drautz2020atomic}. MACE-MP-0 is pre-trained on the MPtrj~\cite{Deng2023} dataset, including a variety of magnetic orders~\cite{batatia2023foundation}. The fitting does not explicitly account for the GGA/GGA+U mixing compatibility corrections~\cite{Wang2021,Jain2011,U-corrections}.
\\
The pre-trained ALIGNN-FF model~\cite{Choudhary2023} is the extension to derivative prediction and force field (FF) formalism~\cite{Choudhary2023} of the Atomistic LIne Graph Neural Network (ALIGNN) models~\cite{Choudhary2021}, able to capture two and three body
interactions and to output more than 70 materials properties~\cite{Choudhary2023}. The ALIGNN~\cite{Choudhary2021} models
were implemented using the deep graph library (DGL)~\cite{wang2019deep} which allows efficient line graph construction and neural message passing. 
As in the case of CHGNet, the message passing occurs both in an atom graph, where atoms are nodes and interatomic bonds are edges, and in a line graph, where bonds are nodes and bond angles are edges~\cite{Choudhary2021}.
For the atom graph, nine input node features are assigned to each node, based on the atomic species involved: electronegativity, group
number, covalent radius, valence electrons, first ionization
energy, electron affinity, block and atomic volume.
ALIGNN is a part of the Joint Automated Repository for Various Integrated Simulations (JARVIS) infrastructure~\cite{Choudhary2020}, a set of databases and tools for materials design. The ALIGNN-FF model was trained on the JARVIS-DFT dataset which contains around 75,000 materials and 4 million energy-force entries,
out of which 307,113 were used in the training~\cite{Choudhary2023}.
\\
For the sake of simplicity, in what follows we use a shorter nomenclature for three out of the four uMLIPs benchmarked: M3GNet-DIRECT~\cite{qi2024robust} will be called M3GNet, MACE-MP-0~\cite{batatia2023foundation} will be referred to as MACE, and ALIGNN-FF~\cite{Choudhary2023} will be termed ALIGNN. 
Some key features of the four uMLIPs are listed in Table~\ref{table:models}.

\begin{table*}[t] 
\centering
\large
\begin{tabular}{l@{\hskip 2em}c@{\hskip 2em}c@{\hskip 2em}c@{\hskip 2em}c@{\hskip 2em}c@{\hskip 2em}c}
\hline
\textbf{uMLIP label} & \textbf{Model} & \textbf{Version} & \textbf{Model Size} & \textbf{Data Set} & \textbf{N$_{\mathrm{{\bf{elements}}}}$}
& \textbf{Data Size} \\
\hline
`CHGNet' & CHGNet~\cite{Deng2023} & 0.3.0~\cite{chgnet-github} & 400K~\cite{Deng2023} & MP-trj~\cite{MP-trj} & 89 & 1.58M~\cite{Deng2023}\\
`M3GNet' & M3GNet-DIRECT~\cite{qi2024robust} & 2021.2.8-DIRECT-PES~\cite{direct-github} &1.1M~\cite{focassio2024performance} & MPF~\cite{Jain2013} + DIRECT~\cite{qi2024robust} 
& 128 & 1.31M~\cite{qi2024robust} \\
`MACE' & MACE-MP-0~\cite{batatia2023foundation} &
2023-12-03-mace-128-L1\_epoch-199~\cite{mace-github} ~\cite{mace-github-another} & 4.7M \cite{deng2024overcoming} & MP-trj~\cite{MP-trj} & 89 & 1.58M~\cite{Deng2023} \\
`ALIGNN' & ALIGNN-FF~\cite{Choudhary2023} & alignnff\textunderscore wt10~\cite{alignn-github} ~\cite{alignn-figshare} & 4.0M ~\cite{alignn-config} & JARVIS-DFT~\cite{Choudhary2020}& 89 &75K~\cite{Choudhary2023,Choudhary2020}\\
\hline
\end{tabular}

\caption{Specifications of the uMLIPs benchmarked. Model Sizes refers to the number of parameters used for training from the different models~\cite{Deng2023,focassio2024performance,deng2024overcoming}. Data Size refers to the number of structures included in the training Data Set~\cite{qi2024robust}. 
N$_{\mathrm{{{elements}}}}$ refers to the number of chemical species (elements) covered.}
\label{table:models}
\end{table*}

\subsection{Equation of state and comparison with all-electron results}
\label{Sec:common_workflow}

\begin{figure*}[!h]
\centering
\includegraphics[width=0.85\linewidth]{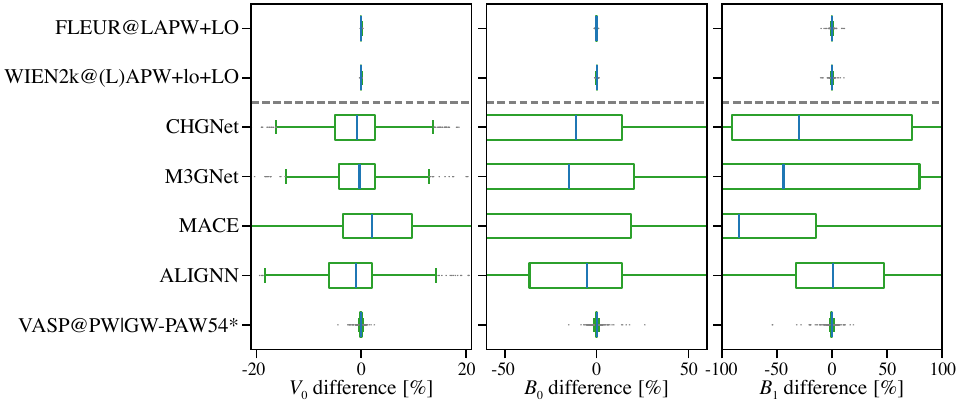}
\caption{
Boxplot showing the relative error in $V_0$, $B_0$ and $B_1$ for the different uMLIPs analyzed in this work and some of the \abinitio codes considered in Ref.~\cite{Bosoni2023} with respect to the average of the AE reference results.}
\label{fig:boxplot_cwf}
\end{figure*}

As a first test, we use the protocol for the equation of state (EOS)
detailed in Ref.~\cite{Bosoni2023}, where a high-quality reference dataset of EOS for 960 cubic crystal structures is generated by employing two all-electron (AE) codes.
This dataset includes all elements from Z = 1 (hydrogen) to Z = 96 (curium).
For each element X, four mono-elemental cubic crystals (\textit{unaries structures}) are considered, in the face-centered cubic (FCC), body-centered cubic (BCC), simple cubic (SC), and diamond crystal structure, respectively.
Besides, for each element, 
six binary cubic oxides (\textit{oxides structures}) are included, with chemical formula X$_2$O, XO, X$_2$O$_3$, XO$_2$, X$_2$O$_5$, and XO$_3$, respectively.
While in Ref.~\cite{Bosoni2023} this dataset is used to gauge the precision and transferability of nine pseudopotential-based \abinitio codes, we use it here to assess the four tested uMLIPs.

In Fig.~\ref{fig:boxplot_cwf}, the relative errors on the uMLIPs predictions for the equilibrium volume $V_0$, the bulk modulus $B_0$, and its derivative with respect to the pressure $B_1$ obtained from a fit~\cite{Bosoni2023} of the EOS,
are compared with 
the analogous errors from selected \abinitio methods employed in Ref.~\cite{Bosoni2023}.
FLEUR~\cite{FLEUR} and WIEN2K~\cite{Blaha2020} are AE codes while VASP~\cite{Kresse1999} implements the 
projector augmented-wave method
(PAW) method~\cite{Blochl1994} with a planewave basis set. The errors reported in Fig.~\ref{fig:boxplot_cwf} are calculated with respect to the average of the AE methods considered here, \ie, FLEUR~\cite{FLEUR} and WIEN2K~\cite{Blaha2020}.

Furthermore, as in Ref.~\cite{Bosoni2023}, the EOS computed with two different computational approaches $a$
and $b$ ($E_a(V)$ and $E_b(V)$) are compared through two metrics $\epsilon(a,b)$ and $\nu(a,b)$.
The first is a renormalized dimensionless version of the metrics $\Delta(a,b)$ introduced in Ref.~\cite{Lejaeghere2016}:
\begin{equation}
\epsilon(a,b) = \sqrt{\frac{\sum_i\left[E_a(V_i)-E_b(V_i)\right]^2}{\sqrt{\sum_i\left[E_a(V_i)-\left<E_a\right>\right]^2\sum_i\left[E_b(V_i)-\left<E_b\right>\right]^2}}} ,
\label{eq:1stmetrics}
\end{equation}
where the index $i$ runs over the explicit calculations of $E_{a,b}(V)$ for the different methods and $\left<E_{a,b}\right>$ is the 
average of $E_{a,b}(V)$ over the considered volume range.
The second metrics, dependent directly on the physically measurable quantities $V_0$, $B_0$, and $B_1$,
captures the relative deviation for each of 
these three parameters between the two computational approaches $a$ and $b$: 
\begin{equation}
\nu_{w_{V_0},w_{B_0},w_{B_1}}(a,b) = 100 \sqrt{\sum_{Y=V_0,B_0,B_1}\left[w_Y\frac{Y_a-Y_b}{(Y_a+Y_b)/2}\right]^2} ,
\label{eq:2ndmetrics}
\end{equation}
where $w_{V_0}$, $w_{B_0}$, and $w_{B_1}$ are appropriately chosen weights (see Ref.~\cite{Bosoni2023}). 
In Eqs.~\eqref{eq:1stmetrics}-\eqref{eq:2ndmetrics}, $a$ is a uMLIP and $b$ is the average of the AE methods considered here, \ie, FLEUR~\cite{FLEUR} and WIEN2K~\cite{Blaha2020}. 
Heatmaps of the periodic table with 
the values of the comparison metrics $\varepsilon$ and $\nu$ obtained with all the different uMLIPs are reported in Supplemental Material Figs.~S1-S4.

It should be noted that most structures in the dataset used in this EOS test are not stable in nature, so that
they were not likely included in the dataset used to train the uMLIPs.
As a consequence, it is not surprising that 
the uMLIPs are not able to predict the correct energy versus volume curve for a significant fraction of the systems (Figs.~S1-S4). 
Nevertheless, even for those systems for which a physical EOS is obtained, the precision and transferability is still far from the one that can be achieved with state-of-the-art pseudopotential-based \abinitio techniques, as can be observed in Fig.~\ref{fig:boxplot_cwf}. 
This is a very stringent test for uMLIPs, especially given the effort made in Ref.~\cite{Bosoni2023} to improve the precision and the transferability of pre-existent pseudopotential tables used for \abinitio calculations.
Yet 
these results suggest that uMLIPs predictions should be taken with some caution and, if possible, validated 
a posteriori via \abinitio calculations, especially if the chemical/physical environment under study is not properly included in the training dataset.
The precision of uMLIPs can be improved after retraining the model by including additional \abinitio data capturing the chemical/physical configurations under investigation~\cite{focassio2024performance, deng2024overcoming}. 
However, despite their relevance, these topics are beyond the scope 
of the present work and are left for future investigation.

\subsection{Structural optimization and formation energies}
\label{Sec:relaxations}

\begin{figure}[!h]
\centering
\includegraphics[width=\linewidth]{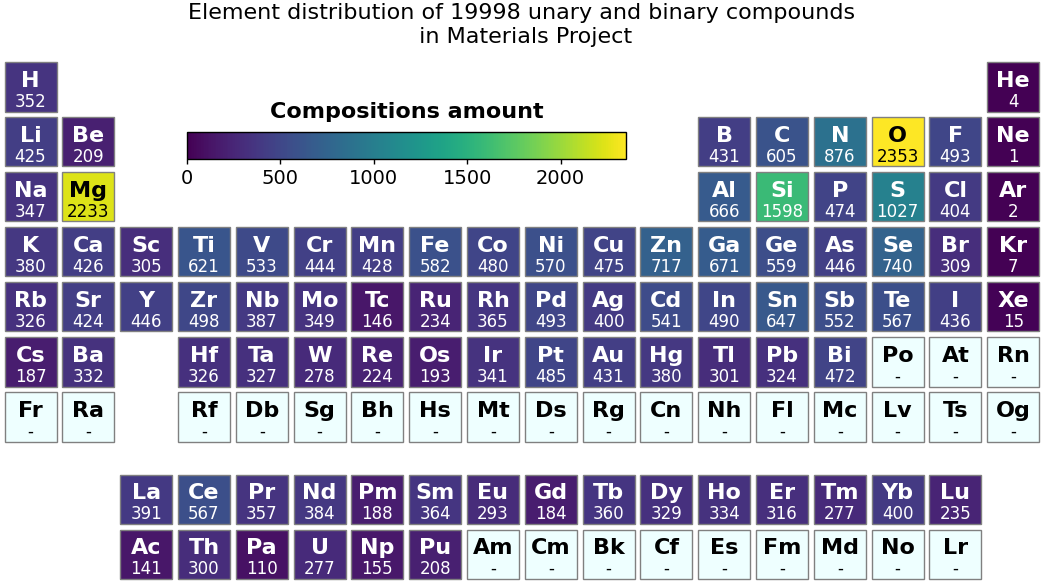}
\caption{
Heatmaps of the periodic table for the dataset chosen for the structural relaxations (unaries and binaries), expressing the data distribution over the chemical composition space. The figure has been produced with the pymatviz tool~\cite{riebesell_pymatviz_2022}.}
\label{fig:Compositions}
\end{figure}

To test the accuracy of the different uMLIPs further, we prepare a dataset with 19998 materials consisting of unary and binary (with 6903 element combinations) phases in MP. It is worthwhile to highlight that the MP database contains structures relaxed with VASP~\cite{Kresse1999} with the PBE functional~\cite{Perdew1996}.
The distribution of the chosen dataset on 
the chemical elements is reported in Fig.~\ref{fig:Compositions}.
We use the different uMLIPs to perform energy calculations both with no structural optimization (we call these ``one-shot'' calculations) and with structural optimization. The latter are of two types: (i) only the atomic positions are relaxed (we call these ``ion-relax'' calculations), and 
(ii) both atomic positions and cell parameters (\ie, lattice parameters and angles) are relaxed (we call these ``cell-relax'' calculations).
For the structural optimizations, in some cases the calculation stops due to errors while building the graph representation ({\em{e.g.,}} given that isolated atoms are found in the structure). Calculations for which this problem appears are tagged as \problematic.
In another non-negligible number of cases, that we tag as \unconverged, the relaxation algorithm is not able to reach the stopping criterion before 150 steps. Finally, we tag as \converged the calculations where the stopping criterion is met in less than 150 steps.
Additional details on the relaxation algorithm are given in Sec.~\ref{Sec:methods}. 

\begin{figure*}[!h]
\centering
\includegraphics[width=0.85\linewidth]{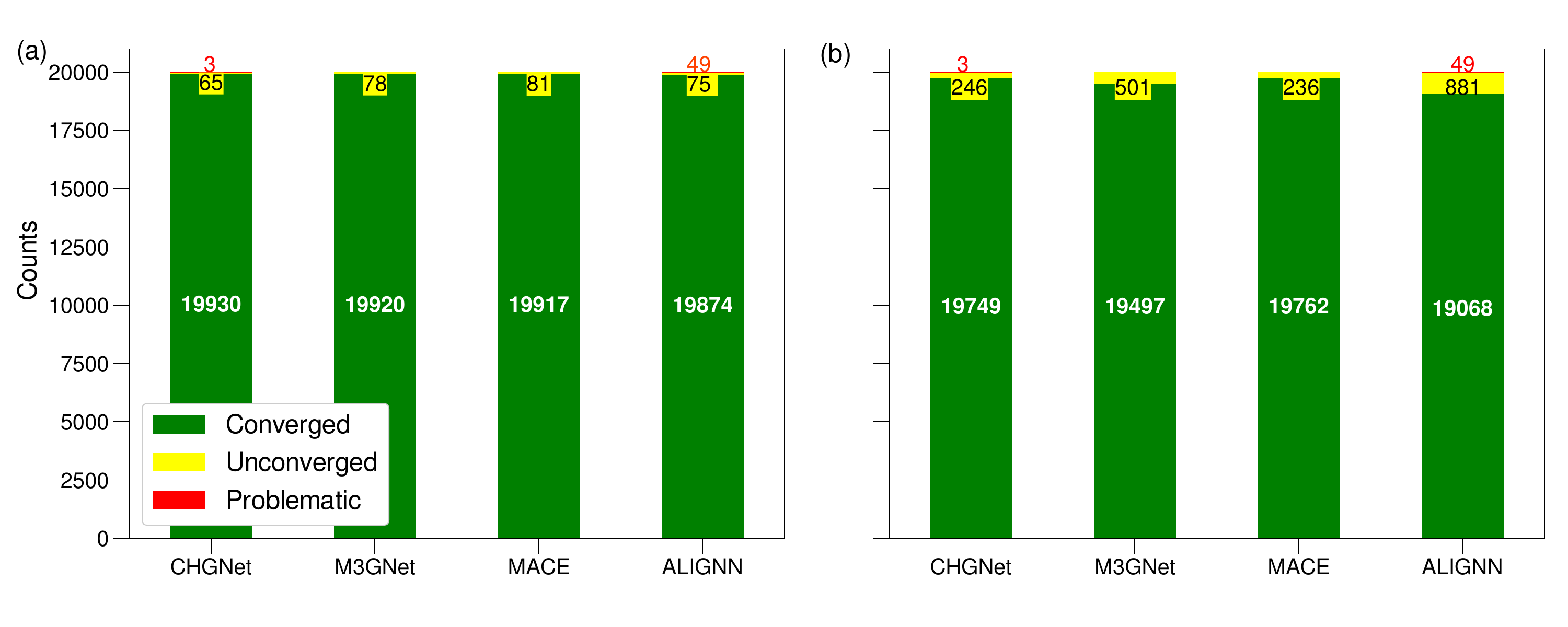}
\caption{
Analysis of the convergence of the different uMLIPs for geometry optimization.
In panel (a), only the atomic positions are relaxed, while the cell parameters are kept fixed at the original values.
In panel (b), both the atomic positions and the cell parameters are relaxed.
As explained in the text, a calculation is considered to be converged when the relaxation criteria (in terms of forces and stresses) are met within 150 steps.}
\label{fig:count}
\end{figure*}

In Fig~\ref{fig:count}, we report the number of occurrences of the \problematic, \unconverged, and \converged tags from the different uMLIPs for the two types of structural optimization.
When only the atomic positions are relaxed (Fig~\ref{fig:count}(a)), the fraction of 
\unconverged cases is very limited (about 0.3$-$0.4\% of the total for all the uMLIPs), while only ALIGNN presents some \problematic cases worth to be noted, but still very small (0.2\%). 
The fraction of \unconverged calculations increases significantly when both atomic positions and cell parameters are optimized (Fig~\ref{fig:count}(b)), especially for M3GNet (2.5\%) and ALIGNN (4.4\%).
CHGNet and MACE perform more 
robustly, with lower fractions of \unconverged results (both 1.2\%). In the following, we discuss only results from the \converged geometry optimizations.

\begin{figure*}
\centering
\includegraphics[width=\textwidth]{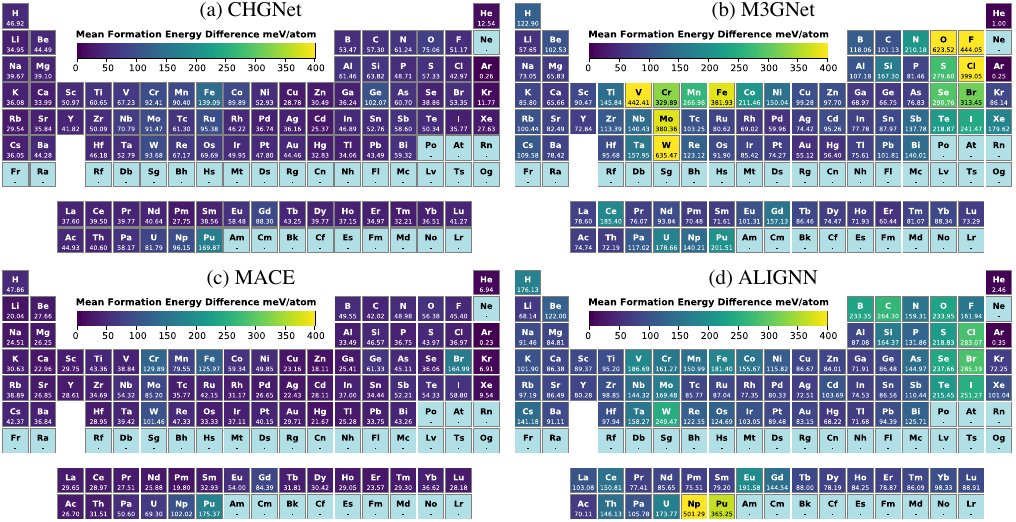}
\caption{Heatmaps of the periodic table for the absolute values of $\Delta E_{\textrm{form}}$ for the one-shot calculations from the different uMLIPS: (a) CHGNet, (b) M3GNet, (c) MACE, and (d) ALIGNN. The figure has been produced with the pymatviz tool~\cite{riebesell_pymatviz_2022}.} 
\label{fig:ptable}
\end{figure*}

\begin{figure}
\centering
\includegraphics[width=\linewidth]{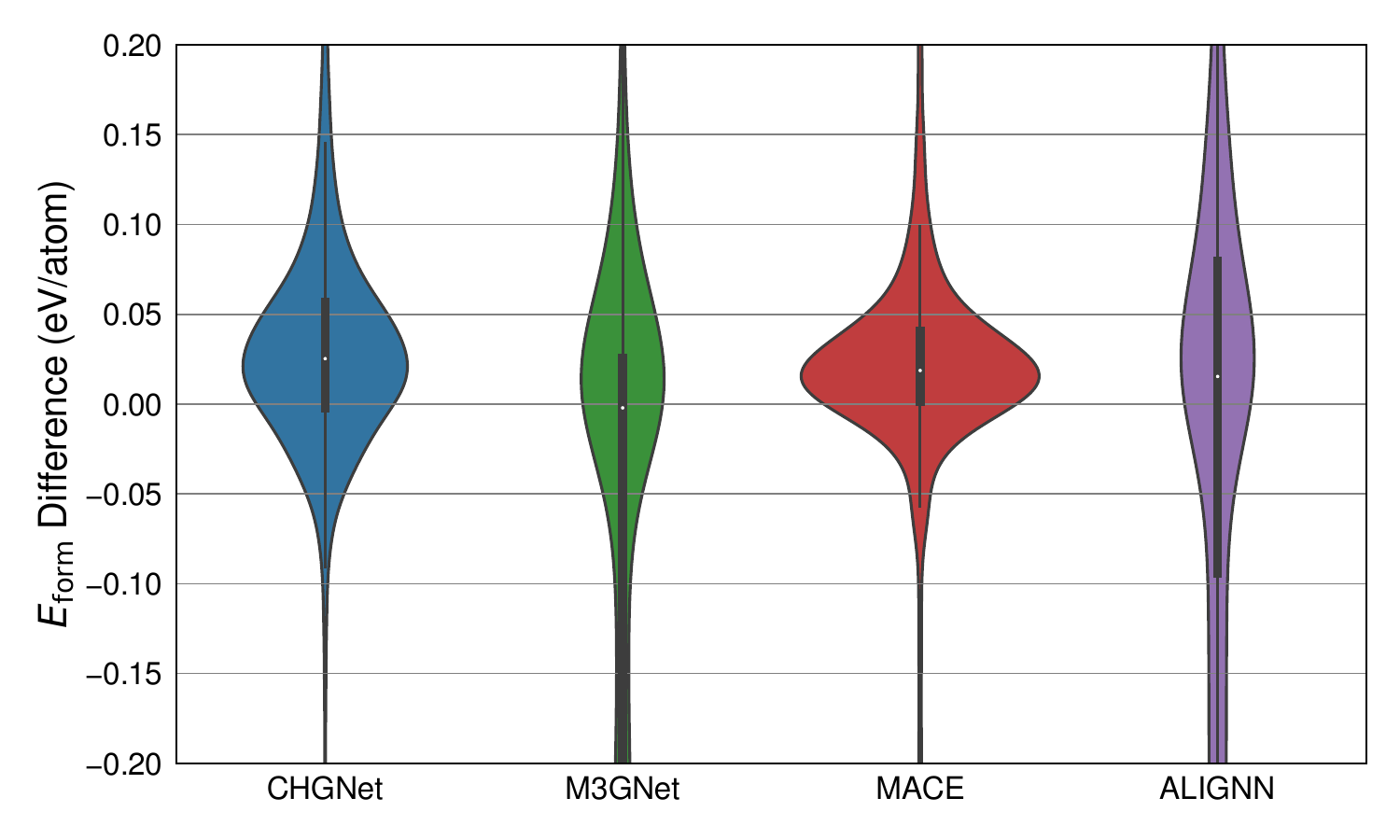}
\caption{Violin plots showing the distribution of $\Delta E_{\textrm{form}}$ (Eq.~\eqref{Eq:eformation}) from one-shot calculations for the different uMLIPs.}
\label{fig:eformviolin}
\end{figure}

To compare the predictive performance of these uMLIPs on the energetics, 
we 
use 
as a benchmark 
the formation energy, defined as
\begin{equation}
E_\textrm{form}[A_aB_b] = E[A_aB_b] - x_a E[A] - x_b E[B],
\label{Eq:eform}
\end{equation} 
where $E[A_aB_b]$ is the total energy per atom of 
the phase of interest with $A$, $B$ the constituent elements, 
$x_a=a/(a+b)$ and $x_b=b/(a+b)$ are the 
fractions of $A$ and $B$, respectively, 
and $E[A]$ and $E[B]$ are the lowest possible energies per atom of 
$A$ and $B$, respectively. The ability of a 
uMLIP 
in predicting formation energies is evaluated 
as a difference from the MP values, \ie, 
\begin{equation}
\Delta E_\textrm{form} = E_\textrm{form}^{\text{MP}} - E_\textrm{form}^{\text{uMLIP}} ,
\label{Eq:eformation}
\end{equation}
with $E_{\textrm{form}}^{\text{MP}}$ and $E_{\textrm{form}}^{\text{uMLIP}}$ being the formation energy 
(Eq.~\eqref{Eq:eform})
from the 
MP and calculated with the uMLIP model, respectively.

We choose to work with the one-shot energies as this allows 
to avoid spurious effects due to the different relaxed geometries predicted by the uMLIPs. 
It is important to note that all our calculations incorporate energy corrections. 
For CHGNet and M3GNet, these corrections 
are already encoded in the uMLIPs, as both models were trained using corrected energies~\cite{Wang2021,Jain2011,U-corrections}. 
Conversely, for MACE, the corrections are applied as a post-processing step (see Sec.~\ref{Sec:methods} and Refs.~\cite{Ong2013,Pymatgen-corr} for details) since this model was trained on uncorrected entries~\cite{batatia2023foundation}.
ALIGNN, on the other hand, 
was trained using the JARVIS-DFT dataset with OptB88vdW functional~\cite{Klimes2009}, and does not use any corrected energies, but rather direct DFT outputs.
In this case, we compute Eq.\eqref{Eq:eformation} with and without energy corrections, and we select the best-agreement set of results.

Figure~\ref{fig:ptable} shows the distribution of the absolute mean values of $\Delta E_\textrm{form}$ on the chemical space for the different uMLIPs, \ie, the average of the absolute value of $\Delta E_\textrm{form}$ over all the chemical systems containing a given element.
CHGNet and MACE have the best performance over all the periodic table. 
For what concerns the elements 
in the transition metal series, CHGNet and MACE perform significantly better 
than M3GNet and ALIGNN, especially for elements like V, Mo, and W. 
In the case of CHGNet, this might be related to the inclusion of the magnetic moments in the training process. CHGNet and MACE outperform M3GNet and ALIGNN also for what concerns the calchogens (O, S, Se, Te) and some halogens (F, Cl, I). ALIGNN is also less performant for actinide elements Np and Pu compared to other uMLIPs.
In Fig.~\ref{fig:eformviolin}, we report the distribution of $\Delta E_\textrm{form}$ (Eq.\eqref{Eq:eformation}) for the one-shot calculations using the different uMLIPs, while the performance of each model
is summarized 
in Table~\ref{table:form_energy_gs}.
MACE shows better performance than the other uMLIPs, with the smallest MAE and RMSE and the highest R\textsuperscript{2} (0.044, 0.101, and 0.989, respectively). 
In turn, MACE outperforms the other uMLIPs in the one-shot calculations (Fig.~\ref{fig:eformviolin} and Table~\ref{table:form_energy_gs}), while displaying, together with CHGNet,
the highest number of 
successful structural optimizations (or \converged results), \ie, 
88.8\%, while M3GNet and ALIGNN display 87.5\% and 85.6\%, respectively (Fig.~\ref{fig:count}).

\begin{figure}[hbt!]
\centering
\includegraphics[width=\linewidth]{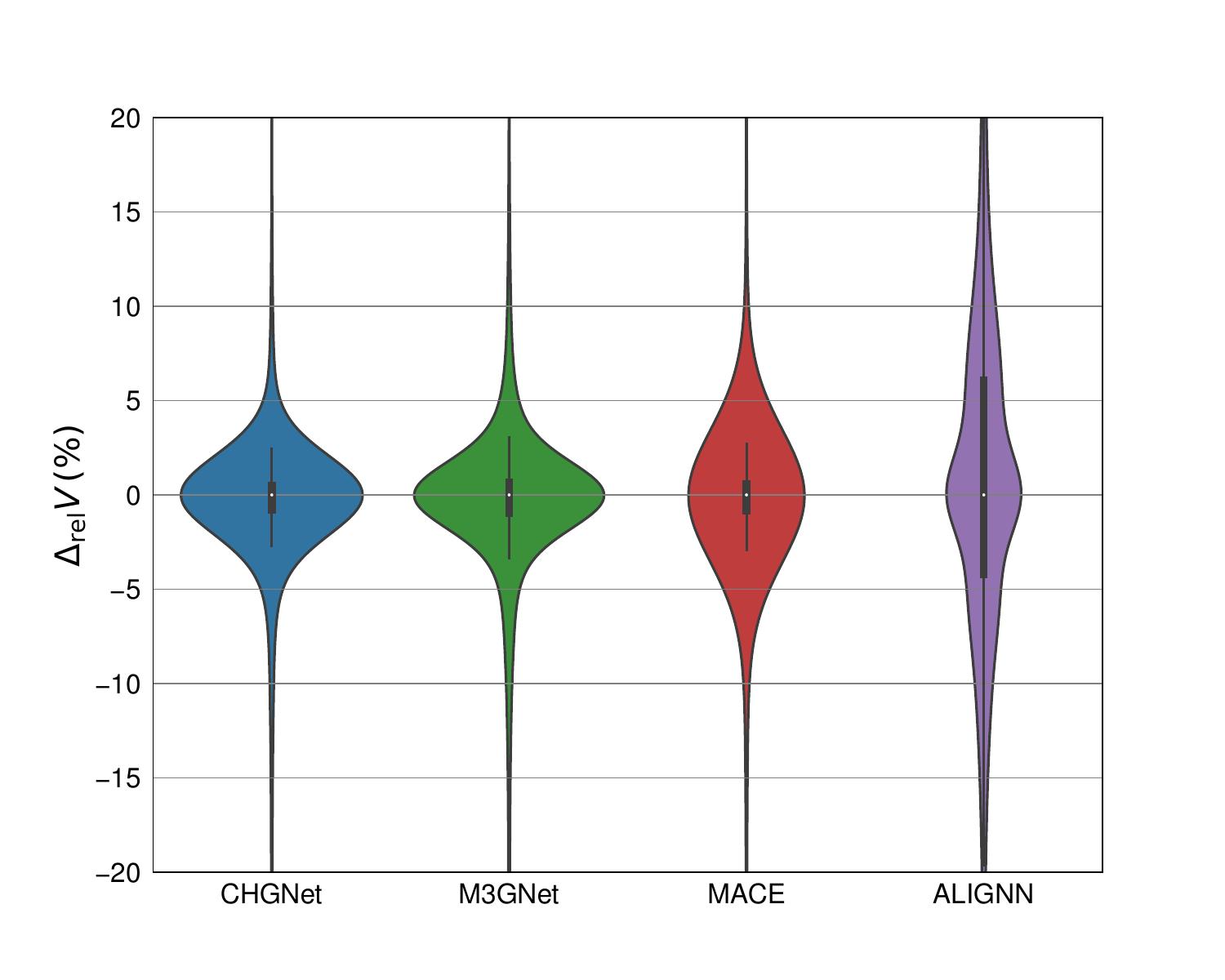}
\caption{Violin plots of the relative difference between the MP cell volume and the one predicted by the different uMLIPs (Eq.~\eqref{eq:volume}). 
Only \converged runs are included in the plot.}
\label{fig:lattice_mae}
\end{figure}

To examine 
the ability of the different uMLIPs 
to predict the cell geometry 
(\ie, lattice parameters, angles, and volume) in the cell-relax calculations, we compare the uMLIPs results with the MP values, as done for the formation energy (Eq.~\eqref{Eq:eformation}). {\em{E.g.}}, for the volume we consider
the relative difference
\begin{equation}
\Delta_\textrm{rel} V = 1 - \frac{V^{\mathrm{uMLIP}}}{V^{\mathrm{MP}}},
\label{eq:volume}
\end{equation}
where $V^{\mathrm{MP}}$ and $V^{\mathrm{uMLIP}}$ are the cell volume from MP and from the uMLIP model cell-relax calculations, respectively. The distribution of $\Delta_\textrm{rel} V$ is reported in Fig.~\ref{fig:lattice_mae}.
CHGNet and M3GNet outperform the other uMLIPs, with a narrower distribution of $\Delta_\textrm{rel} V$. 
ALIGNN performs the worst, while MACE has an intermediate performance.
In Table~\ref{table:lattice}, we 
report the Mean Absolute Relative Error (MARE) on the predicted lattice parameters, angles, and volumes. We observe that MACE shows similar performance with respect to CHGNet and M3GNet 
for $a$, $b$, $c$ and $\alpha$, $\beta$, $\gamma$, but its performance on the volume is 
significantly worse than the other two uMLIPs, as also shown in Fig.~\ref{fig:lattice_mae}. 
One possible origin of this behavior might be the compensation of the errors of the cell parameters in the case of M3GNet and CHGNet, which would not occur in MACE.
Another reason could be the presence of more outliers (extreme large deviations) in the case of MACE, as documented from the histograms of the relative volume differences and lattice parameters differences provided in the Supplemental Material.

A similar 
argument can be applied to the slightly higher MARE on the volume for CHGNet as compared to M3GNet (Table~\ref{table:lattice}). 
ALIGNN shows larger errors compared to the other uMLIPs (Table~\ref{table:lattice}).
To further test the ability of uMLIPs to treat multi-element compounds, we perform `one-shot' and `cell-relax' calculations on 100 randomly chosen quinary materials.
There are 4 \unconverged cases (4\%) for CHGNet and M3GNet, while 2 \unconverged cases (2\%) for MACE and ALIGNN.
This seems to indicate that these uMLIPs can still be used for structural relaxations even for quinaries 
although additional investigations would be needed to assess 
the transferability in this part of the chemical space. 
In the Supplemental Material we report the comparison of formation energies and cell geometries of quinaries with the respective MP reference data, as done for the unary and binary compositions (Figs.~\ref{fig:eformviolin} and \ref{fig:lattice_mae}). 
This comparison indicates that MACE still gives the most reliable prediction on formation energies, 
while CHGNet shows the best 
performance for cell geometry 
predictions.

\begin{table}[ht!]
\centering
\begin{tabular}{l@{\hskip 2em}c@{\hskip 2em}c@{\hskip 2em}c}
\hline
\textbf{uMLIP} & \textbf{MAE} & \textbf{RMSE} & \textbf{R\textsuperscript{2}} \\
\hline
CHGNet & 0.054 & 0.105 & 0.988 \\
M3GNet & 0.172 & 0.316 & 0.896 \\
MACE & 0.044 & 0.101 & 0.989 \\
ALIGNN & 0.137 & 0.2226 & 0.947\\
\hline
\end{tabular}
\caption{Performance of the uMLIPs in predicting the formation energy for one-shot calculations. As in Fig.~\ref{fig:eformviolin} for the energies, here MAE and RMSE are in eV/atom.}
\label{table:form_energy_gs}
\end{table}

\begin{table}[ht!]
\centering
\begin{tabular}{l@{\hskip 1.5em}r@{\hskip 1.5em}r@{\hskip 1.5em}r@{\hskip 1.5em}r@{\hskip 1.5em}r@{\hskip 1.5em}r@{\hskip 1.5em}r}
\hline
\textbf{uMLIP} & volume & a & b & c & $\alpha$ & $\beta$ & $\gamma$ \\
\hline
CHGNet & 3.16 & 2.03 & 2.07 & 2.44 & 0.75 & 0.62 & 1.19 \\
M3GNet & 2.97 & 2.04 & 2.09 & 2.46 & 0.89 & 0.73 & 1.24 \\
MACE & 5.22 & 2.01 & 2.11 & 2.58 & 0.73 & 0.59 & 1.13 \\
ALIGNN & 7.85 & 3.42 & 3.42 & 3.61 & 0.94 & 0.86 & 1.32 \\
\hline
\end{tabular}
\caption{
Mean Absolute Relative Error (MARE in \%) of the different uMLIPs in predicting the volume, lattice parameters, 
and angles.} 
\label{table:lattice}
\end{table}

\subsection{\label{Sec:phonons} Vibrational properties}

In this section we analyze the capability of uMLIPs to reproduce the vibrational properties of crystalline materials using accurate \abinitio results reported in a previous work~\cite{Petretto2018a} as a reference. From the structures in Ref.~\cite{Petretto2018a} we select those whose energy above hull is zero
both in the MP database and from the uMLIP cell-relax calculations, leading to 101 structures. Phonons are computed using the finite displacement method as implemented in 
the \phonopy package~\cite{phonopy-phono3py-JPCM,phonopy-phono3py-JPSJ}. 
Further details on the protocol employed to compute phonons with uMLIPs are provided in Section~\ref{Sec:methods}.
By exploiting the finite displacement method to compute phonons, this study indirectly probes the quality of the uMLIPs-calculated forces when atoms are slightly displaced away from the equilibrium positions. 
From a methodological point of view we note that, already when \abinitio engines are used, the accuracy of phonon calculations from the finite displacement method is rather sensitive to the quality of the force 
calculations in the supercell.
Since by construction these forces are less accurate when calculated from uMLIP models than from \abinitio methods, it is reasonable to expect some non-negligible discrepancy between uMLIPs and \abinitio phonon calculations. 
Also, as discussed in more detail in Sec.~\ref{Sec:methods}, the MLIP architectures at the basis of the universal models considered in this study cannot predict
the long-range dipolar contributions to the interatomic force constants, 
needed to obtain a reliable Fourier phonon interpolation in the case of polar materials~\cite{Gonze1997a, Baroni2001}. The uMLIP-calculated forces should therefore be augmented with the ML electronic dielectric tensor and the Born effective charges, which 
are not available at present, leading us to augment the uMLIPs forces with \abinitio~\cite{Gonze2020,Romero2020} values for these quantities 
(see Sec.~\ref{Sec:methods} for further details).

\begin{figure}
\centering
\includegraphics[width=\linewidth]{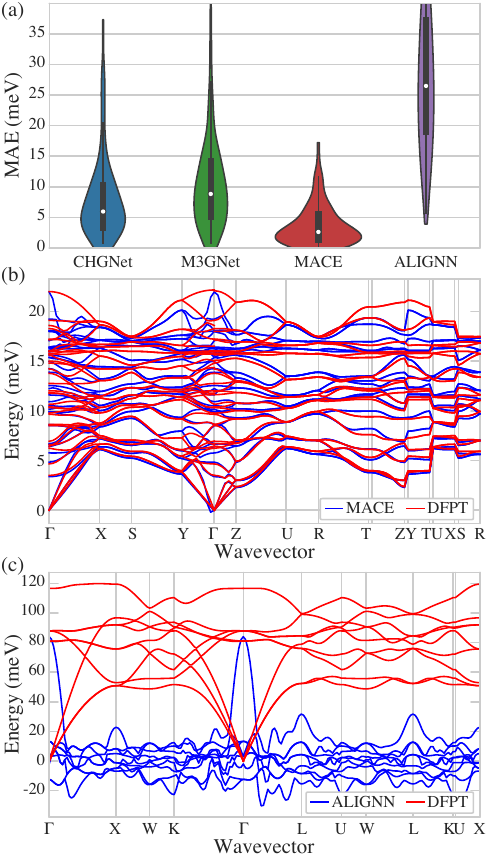}
\caption{(a) Violins plots of the MAE (Eq.~\eqref{eq:phonons}) on the computed phonon band structures from uMLIPs and DFPT.
(b) Comparison of the phonon band structures computed with DFPT and
MACE for the compound (mp-567744: SrBr$_2$) with the smallest MAE (0.3~meV).
(c) Comparison of the phonon band structures computed with DFPT and
ALIGNN for the compound (mp-1569: Be$_2$C$_1$) with the largest MAE (75.4~meV).}
\label{fig:phonons}
\end{figure}

For the above 
reasons, when comparing uMLIPs results with \abinitio data, we choose a rather generous metrics, \ie, the MAE between the uMLIPs and the \abinitio phonon band structures computed along a high-symmetry $\qq$-path~\cite{Setyawan2010}:
\begin{equation}
\text{MAE} = \dfrac{1}{N_\qq}\sum_{\qq\nu} |\wqnu^{\text{uMLIP}} - \wqnu^{\text{DFPT}}|.
\label{eq:phonons}
\end{equation}
In Eq.~\eqref{eq:phonons}, $\omega_{\qq\nu}$ is the phonon energy in meV, $\nu$ is the branch index, $N_\qq$ is the number of wavevectors used to sample the $\qq$-path, and $\omega_{\qq\nu}^{\text{uMLIP}}$, $\omega_{\qq\nu}^{\text{DFPT}}$ are the phonon energies computed using the uMLIP model and the density-functional perturbation theory (DFPT) with ABINIT~\cite{Petretto2018a}, respectively.
Table~\ref{table:min_mae_avg_phonon} reports the minimum, maximum, and average values of the MAE in the phonon band structures obtained from the different uMLIPs (Eq.~\eqref{eq:phonons}),
while the MAE distribution 
is given in
panel (a) of Fig.~\ref{fig:phonons}.

\begin{table}[ht!]
\centering
\begin{tabular}{lccc}
\hline

\textbf{uMLIP} & \textbf{MIN\_MAE} & \textbf{MAX\_MAE} & \textbf{MEAN\_MAE} 
\\ 
\hline
CHGNet & 0.82 & 37.34 & 8.12 \\
M3GNet & 0.74 & 40.20 & 10.41 \\ 
MACE   & 0.31 & 17.22 & 3.71 \\
ALIGNN & 5.60 & 75.38 & 29.36 \\
\hline
\end{tabular}
\caption{Minimum, maximum and average MAE in meV for the phonon band structures computed from 
different uMLIPs.}
\label{table:min_mae_avg_phonon}
\end{table}

From panel (a) of Fig.~\ref{fig:phonons} and Table~
\ref{table:min_mae_avg_phonon}, 
we observe a non-ideal agreement between uMLIPs-calculated and \abinitio-calculated vibrational properties. 
In addition to the fundamental reasons discussed above, in this particular study a source of discrepancy could come from the fact that Ref.~\cite{Petretto2018a} exploits the PBEsol exchange-correlation functional while the uMLIPs have been trained using PBE~\cite{Perdew1996} data with the inclusion of U-corrections~\cite{Wang2021,Jain2011,U-corrections} for certain systems (in the case of CHGNet and M3GNet) or PBE~\cite{Perdew1996} data without the inclusion of U-correction (in the case of MACE) or optB88vdW~\cite{Klimes2009} data (in the case of ALIGNN).
In panel (b) of Fig.~\ref{fig:phonons}, we report the phonon band structure for the system with the lowest MAE (0.312 meV, obtained from MACE, Table~\ref{table:min_mae_avg_phonon}), which is Sr$_4$Br$_8$ (mp-567744). 
Conversely, in panel (c) of Fig.~\ref{fig:phonons}, we report the phonon band structure for the system with the highest MAE (75.375 meV, obtained from ALIGNN, Table~\ref{table:min_mae_avg_phonon}), which is Be$_2$C (mp-1569).
For the latter, the uMLIP predicts vibrational instabilities that are not
observed in the \abinitio results. However, we also note that even for the lowest-MAE material (panel (b) of Fig.~\ref{fig:phonons}), the high-energy region of the phonon band structure is still far from being accurate. 

To summarize, our results indicate that presently available uMLIPs can predict 
\abinitio vibrational properties with a typical error of 3.71 meV in the best-case scenario.
This value should be considered as a lower bound as the discrepancy is expected to increase 
if the electronic dielectric tensor and the Born effective charges were predicted with ML techniques.
In the opinion of the authors, the predictive behavior of uMLIPs for vibrational properties can be improved by training new uMLIPs with larger weights for the forces loss function but it is also clear that for accurate ML-based predictions in polar materials, one needs uMLIPs capable of inferring the long-range part of the dynamical matrix.
All this being said, we believe that uMLIPs represent 
an efficient and promising approach to perform an initial screening for vibrational and thermodynamic properties, especially in a high-throughput context in which high accuracy is not necessarily needed.

\section{Conclusions}\label{Sec:conclusions}

We present a systematic assessment of various universal machine-learning
interatomic potentials (uMLIPs) by investigating their capability to reproduce
\abinitio results for several important physical properties such as equation of
state, relaxed geometries, formation energies, and vibrational properties of an
extensive set of crystalline materials.
Among the considered uMLIPs, we find that MACE shows superior accuracy in
predicting formation energies and vibrational properties, and CHGNet and M3GNet 
are outstanding for relaxed geometry predictions.
MACE and CHGNet show superior performance on the formation energy prediction
along the periodic table, especially when considering systems containing
transition metal elements, chalcogens, and halogens (with Br being accurately
reproduced only by CHGNet).
M3GNet exhibits relatively high errors, leading to a lower R\textsuperscript{2}
value compared to the other uMLIPs in predicting formation energies.
Despite this, its ability to accurately predict volume and lattice parameters
remains high, and it has intermediate performance in predicting vibrational
properties.
For what concerns ALIGNN, despite trained with different dataset, it still shows
relatively good results when predicting formation energies.
However, it also reveals to be the most problematic model when performing
geometry optimization at variable cell and for predicting vibrational
properties.
These results underscore the need for further optimization and training of the
currently available uMLIPs to fully exploit the capability of ML techniques
across a broader range of applications.
The choice of a particular uMLIP for specific applications should take into
account an appropriate balance between accuracy and computational efficiency.
Future work should aim at enhancing the performance of these potentials further,
particularly focusing on areas where current uMLIPs exhibit limitations such as 
more accurate prediction of forces and stresses or the capability of learning
Born effective charges and electronic dielectric tensors that are crucial for
the vibrational properties of polar materials.
Our work will hopefully pave the way towards a more systematic assessment of
uMLIPs in different scenarios and the establishment of a standardized benchmark
set that can be used to gauge the precision and transferability of uMLIPs.

\section{Methods}\label{Sec:methods}

The calculations are performed with the \abipy
package~\cite{Gonze2020,Romero2020}, more specifically the \texttt{abiml.py}
script providing a unified interface that allows one to perform different types
of calculations such as structural relaxations, molecular dynamics, or NEB using
the algorithms implemented in ASE~\cite{HjorthLarsen2017} and different uMLIPs
as calculators.
The following package versions are used to produce the results reported in this
work.
Python: 3.11, Pymatgen: 2023.7.17, \abipy: 0.9.6, \abinit: 9.8.4, CHGNet: 0.3.2,
ALIGNN-FF: alignn 2023.10.1, MatGL: 1.0.0, MACE-MP0: mace-torch 0.3.4, and ASE:
3.22.1.

Structural relaxations are performed using the ASE optimizer, employing the BFGS
algorithm.
To ensure good trade-off between accuracy and computational cost, the stopping
criterion fmax is set to 0.1 eV/\AA.
If the stopping criterion is not met at a maximum number of iterations set to
150, the calculation is considered unconverged.
This is justified by the fact that all the initial structures in our dataset are
taken from the Materials Project~\cite{Jain2013}, where they have been already
relaxed with VASP~\cite{Kresse1999} with the PBE functional~\cite{Perdew1996},
and the same structures are supposed to be included in the set used to train the
uMLIPs (with the exception of ALIGNN).
It should be pointed out that for each composition the total energy obtained
from the Materials Project~\cite{Jain2013} is obtained with energy
corrections~\cite{Wang2021,Jain2011,U-corrections}.
Therefore, we apply the same correction according to the methods in
\texttt{compatibility.py} from the Pymatgen package~\cite{Ong2013,Pymatgen-corr}
to MACE since it was trained with Materials Project uncorrected energies.

For the phonon calculations we first perform a structural relaxation with the
uMLIPs from the crystalline structures employed in~\cite{Petretto2018a}.
The atomic positions are relaxed at fixed cell parameters in order to avoid
spurious effects due to the change of the lattice parameters with respect to the
reference DFT results.
The relaxed configuration is then used to compute vibrational properties using
the finite displacement method implemented in
\phonopy~\cite{phonopy-phono3py-JPCM,phonopy-phono3py-JPSJ} with a displacement
of 0.01 \AA.
In each calculation, the real-space supercell is matched with the $\qq$-mesh
employed in~\cite{Petretto2018a} to compute the dynamical matrix with the DFPT
part of \abinit~\cite{Gonze2020,Romero2020}.
It is noteworthy that the uMLIPs employed in this work lack the capability to
predict the electronic dielectric tensor $\epsilon^\infty$ and the Born
effective charge tensor $\Zstar_{\kappa}$ where $\kappa$ is the index of the
atom in the unit cell.
As discussed in~\cite{Gonze1997a, Baroni2001}, these quantities are needed to
model the long-range dipolar contribution to the interatomic force constants in
polar materials.
This term is indeed responsible for the LO-TO splitting as well as for the
non-analytical behaviour of the vibrational spectrum near $\textbf{q}=0$.
Its correct numerical treatment is therefore crucial to obtain an accurate
Fourier interpolation of the dynamical matrix at arbitrary $\qq$-points.
For this reason, in all \phonopy calculations we use the \abinitio values of
$\epsilon^\infty$ and $\Zstar_{\kappa}$ obtained with \abinit to model
long-range interactions.
The \abinitio phonon dispersions is obtained by using the \anaddb
post-processing tool using the DDB files retrieved from the MP database to
compare the different phonon dispersions between uMLIPs and \abinitio DFPT
results.
\\
Figure~\ref{fig:boxplot_cwf} and the heatmaps of the periodic table in the
Supplemental Material have been produced using the \abinitio results and the
open source python scripts available in 
the~\href{https://github.com/aiidateam/acwf-verification-scripts}{acwf-verification-scripts}
github repository.

\bmsection*{Author contributions}

G.-M. R. designed the research topic and coordinated the whole project.
All authors provided the ideas underlying this work, contributed to its
development, and discussed the ﬁndings reported in the paper.
All authors took part to the writing and reviewing of the paper, and to the ﬁnal
approval of its completed version.
H. Y. and M. G. executed the calculations presented.
M. G.  developed and maintained the software infrastructure.

\bmsection*{Acknowledgments}

The authors acknowledge useful discussions about the correct use of the different uMLIPs with Yuan Chiang, Kamal Choudhary, Bowen Deng, Tsz Wai Ko, and Shyue Ping Ong.
Computational resources have been provided by the supercomputing facilities
of the Université catholique de Louvain (CISM/UCL),
and the Consortium des Equipements de Calcul Intensif en
Fédération Wallonie Bruxelles (CECI). This work is supported by the National Key Research and Development Program of China(2022YFE0141100 and 2023YFB3003005).

\bmsection*{Financial disclosure}

None reported.

\bmsection{Conflict of interest}

The authors declare no potential conflict of interests.

\bibliography{references.bib}

\bmsection*{Supporting information}

Additional supporting information may be found in the
online version of the article at the publisher’s website.

\end{document}